\documentstyle{elsart}

\def\beq{\begin{equation}}
\def\eeq{\end{equation}}

\def\bea{\begin{eqnarray}}
\def\eea{\end{eqnarray}}
\def\ba{\begin{array}}                  
\def\ea{\end{array}}

\begin{document}
\begin{frontmatter}
\title{On the Gravitational Energy Shift for matter waves}

\author{M. Martinis and N. Perkovi\'c  }
\address{Theoretical Physics  Division, Rudjer Bo\v skovi\' c Institute, Zagreb, Croatia}

\begin{abstract}
The gravitational energy shift for photons  is extended to all
mass-equivalent energies $E = mc^2$, obeying the quantum condition
$E = h\nu$.On an example of a relativistic binary system, it was
 shown that  the  gravitational energy shift would
 imply,in contrast to Newtonian gravity, the  gravitational
 attraction    between  full mass-equivalent energies.
 The corresponding space-time metric becomes exponential.
  A good  agreement was found with all results of weak
field tests of General relativity. The strong field
effects in a  binary system can be easily studied.  A long
standing problems of Pioneer and other flyby anomalies were also
discussed in connection with the violation
of total energy conservation. It was shown that
relatively small energy non-conservation
during the  change  of the orbit  type   could
 explain these persistent anomalies.

\end{abstract}
\end{frontmatter}

\section{Introduction}

The gravitational energy shift for photons  is now a well
established and experimentally tested phenomena\cite{Ripka},
predicted by  Einstein's Equivalence principle \cite{Einstein}.
However, Einstein also established  the well known mass-energy
relation , $E = mc^2$, according to which all types of energies
should be subjected to a gravitational energy shift.  The
corresponding gravitational frequency shift for all matter waves is
then obtained by making use of  the quantum condition$E = h\nu$.
This extension has some interesting consequences worth exploring
here.

As a particular example ,let us   consider   an isolated, compact
self-gravitating binary system with point-like proper masses $M_0$
and $m_0 << M_0$. In the rest frame  of $M_0$, the gravitational
field around $M_0$ is approximately static, spherically symmetric,
and isotropic. The motion of the mass $m_0$    is then
 described by the  following Lagrangian:
\begin{equation}
L  = - m_0c^2\frac{d\tau}{dt},
\end{equation}

 where
 \begin{equation}
 d\tau =
dt\sqrt{g_{tt}(r)- g_{rr}(r)\vec{\beta}^2} = dt_g\sqrt{1 -
\vec{\beta}_g^2}
\end {equation}
  links together three different clocks showing:
  the   proper time, $\tau$, the observer's time,
 $t$, and the gravitational time, $t_g$, respectively.
  The components of
  space-time metric different from zero are $g_{tt}$ and
 $g_{rr} = g^{tt}$\cite{Rudmin}. Different  $\vec{\beta}$-symbols
  denote    dimensionless velocities,
 $\vec{\beta} = \vec{v}/c$ and $\vec{\beta}_g =
 g_{rr}\vec{\beta}$, respectively.
  Note that $c\vec{\beta}_g = d\vec{r}_g/dt_g$
 where $d\vec{r}_g = \sqrt{g_{rr}}d\vec{r}$ and
 $dt_g = \sqrt{g_{tt}}dt$.

  In the  4-vector  notation,  $p^{\mu} = (E/c, p^i)$,
   the relativistic energy $E$, and the 3-momenta $p^i$
   of the mass $m_0$
   are defined as: $ \
  E = mc^2 =m_0c^2dt/d\tau$ and $
  p^i = mv^i = m_0v^idt/d\tau$.
  .
  The corresponding canonical 4-momenta, $p_{\mu} =(H/c,p_i)$
  are obtained from the Lagrangian using:
\begin{equation}
 p_i  =  \frac{\partial L}{\partial v^i}
  =  g_{rr}p^i,
\end{equation}

and $ H  =   p_iv^i - L$  to  find:

\begin{equation}
 H = g_{tt}E =\sqrt{g_{tt}}\sqrt{p_ip^ic^2 + m_0^2c^4}.
\end{equation}

  The orbits of  $m_0$ are characterized by two constants
  (integrals)    of motion: the total energy,$H$ and
  the orbital angular
  momentum,   which   are both derived from the Lagrangian
  equation of motion, or more directly
  from the equivalent mass-shell   condition: :
  \begin{equation}
  HE = p_ip^ic^2 + m_0^2c^4.
  \end{equation}

  In the  polar coordinate system,
   ($r, \theta,\varphi$), the equatorial motion
   ( $ \theta = \pi/2$),
    of   the mass $m_0$ is described by:

    \begin{equation}
    \Big(\frac{dr}{d\tau}\Big)^2 =
    c^2(h^2 - g_{tt}-    g_{tt}^2\frac{l_{\varphi}^2}{r^2})
    \end{equation}

    where $h = H/m_0c^2$ and
    $cg_{tt}l_{\varphi} = r^2d\varphi/d\tau$ are two integrals
    of motion   expressing the conservation of total
    energy and angular momentum.
    Stable  bound and unbound orbits are determined  by
    solving the     algebraic equation:
    \begin{equation}
    h^2 = g_{tt}+
    g_{tt}^2\frac{l_{\varphi}^2}{r^2}.
    \end{equation}

     For a given $g_{tt}$, the strong field orbital
    characteristics  are  best
    studied in the  ($ u,\varphi$) coordinates , where
    $u = R_g/r$ and
    $R_g =  GM_0/c^2$ denotes
     the gravitational radius of $M_0$.
     In these coordinates (6) becomes

    \begin{equation}
    \Big(\frac{du}{d\varphi}\Big)^2 + u^2 =
    (h^2g_{rr} - 1)\frac{g_{rr}}{l^2}
    \end{equation}

    where $R_gl = l_{\varphi}$. .

    In the next Section, we are going to show how the gravitational
  energy shift   can be used to obtain
  an explicit  form for the  metric $g_{tt}(r)$.

\section{Gravitational energy shift for matter waves}

  The total gravitational energy shift consists of two different
  contributions: one due to
  the position
  of the mass $m_0$ in  the gravitational field,  and other due to its
    relative motion. This becomes obvious if we
  express the relativistic energy of the mass$m_0$ in the
   form :

\begin{equation}
 E =\frac{E_g}{\sqrt{1 - \vec{\beta}_g^2}},
\end{equation}

 where $E_g = m_0c^2/\sqrt{g_{tt}}$ and $\vec{\beta}_g =
 g_{rr} \vec{\beta}$.

  Then, the  total  gravitational energy shift, defined as
  $\Delta E/E$  is
  \begin{equation}
  \frac{\Delta  E}{E} = \frac{\Delta  E_g}{E_g} +
  \frac{\vec{\beta}_g\Delta \vec{\beta}_g}{1 - \vec{\beta}_g^2},
  \end{equation}

  where $\Delta  E_g/E_g$ denotes a pure gravitational energy shift
  defined as a ratio between
  $E_g$-energies at two
different radial distances of the mass$m_0$ from $M_0$ :
\begin{equation}
\frac{E_g(r_2)}{E_g(r_1)}=  \frac{\Delta E_g}{E_g} + 1 =
\sqrt{\frac{g_{tt}(r_1)}{g_{tt}(r_2)}}.
\end{equation}

In the weak field approximation,  $ g_{tt} = 1 - 2u + ...$, a well
known expression
\begin{equation}
\frac{\Delta E_g}{E_g} = \frac{R_g}{r_1r_2}(r_1 - r_2)+ ....
\end{equation}
is obtained.  The full gravitational frequency  shift  is obtained
by replacing $E = E_g/\sqrt{1 = \beta_g^2}$ with $h\nu$ and $E_g$
with $h\nu_g$.
 The   higher order terms in (12)  become important at
 relativistic velocities, $\beta_g \sim 1$, and
 in  strong gravitational fields , when
  an explicit form of  $g_{tt}$ is needed.
  The Einstein's general relativity offers one  solution,
   the Schwarzschild metric \cite{Weinb}, but there are many
alternative solutions that   all give the same  weak-field results
\cite{PNP}.

However, we have found one more or less obvious solution for
$g_{tt}$, based on the infinitesimal form,  of the  gravitational
energy shift,$dE_g/E_g = d(R_g/r) = du$. The formal integration of
this equation yields to $E_g = m_0c^2exp(u)$ and to
\begin{equation}
g_{tt} = e^{- 2u} .
\end{equation}
We recognize here  the exponential metric, first introduced by
Yilmaz \cite{Yilmaz}.

 The most interesting feature of this solution is the observation
 that  the gravitational energy shift
  also implies the form  of the radial
gravitational force which is according  to
\begin{equation}
dE_g = - dU_g = drF_g = E_g du,
\end{equation}
where $U_g = m_0c^2(1 - e^u)$ is the c gravitational potential
energy of the form
\begin{equation}
F_g = -Ge^u\frac{M_0m_0}{r^2}.
\end{equation}
  It modifies   Newtonian gravity at short distances by
allowing  for a variation  of the gravitational constant $G$ with
distance, $G(r) = G e^u$ \cite{Moffat}. This radial  force can also
be interpreted as a force acting between two full mass-equivalent
energies, in our case $M_0c^2$ and $ E_g =  m_0e^uc^2$. In the following,
 we shall apply this new force to  the problem of explaining  the Pioneer
and flyby anomalies.

 \section{On the Pioneer and Flyby Anomalies}

The Pioneer anomaly refers to a systematic  observation
\cite{Anders1} that NASA's two Pioneer 10 and 11
spacecrafts,launched in the early 1970s,  are slowing down  more
than expected on their ways  out of the solar system.
 This anomalous  deceleration  towards
 the Sun  is almost constant and uniform  of a magnitude
 $a_P = (8.74 ± 1.33)\times10^{-10}$ ms$^{-2}$, for
heliocentric distances greater than 20 AU \cite{Anders1}. This
anomaly remained largely unexplained within the currently accepted
Newton-Einstein laws of gravitation \cite{Iorio}. Any pure
gravitational explanation of the Pioneer anomaly  should also face a
a  difficult problem: of  agreeing  with the  very precise
 cartographic data  of the solar system. Similar temporal
 anomalies  have  been observed during several planetary flyby
 missions \cite{Anders3}. It is interesting that all anomalous
 accelerations occurred after the spacecrafts have changed  their orbital
 parameters  by means of planetary flybys, during which the
 energy can not be conserved. This temporal energy transfer
 can be studied within our
relativistic binary model applied to the Sun (Earth)-spacecraft
system. During the flyby energy transfer the spacecraft radial
acceleration consists of two contributions:one coming from the
energy conserving part and thr other coming  from  the energy
non-conserving part, in the form $a = a_C + a_A$
 where
\begin{eqnarray}
a_C & = & \frac{1}{2}\frac{\partial}{\partial
r}\Big(\frac{dr}{dt}\Big)^2 \\
 a_A & = &
\frac{1}{2}(\frac{dh}{dr})\frac{\partial}{\partial
h}\Big(\frac{dr}{dt}\Big)^2\
    \end{eqnarray}

 where connection with $\vec{v}^2$ is given by\\
 $\vec{v}^2 = c^2(1 - \frac{g_{tt}}{h^2})g_{tt}^2
   = \Big(\frac{dr}{dt}\Big)^2 +
 c^2g_{rr}^4(\frac{l_{\varphi}}{hr})^2$.\\
  The energy transfer during the  change of the  orbit type
  is accompanied with a temporal non-conservation of total energy , it is not difficult
 $h$, resulting in an anomalous acceleration,
 \begin{equation}
2a_A = - c^2[1 - 6u + (6\beta + 12 + l^2)u^2 +
...]\frac{d}{dr}(h^{-2}),
\end{equation}
where we  have used
 $g_{tt} = 1 -2u +2\beta u^2 + O(u^3)$. The
  exponential metric is characterized by $\beta = 1$.
  In the case of a Pioneer anomaly the required
  relative violation   of  total energy  is of
  the order of $10^{-16}$, or more precisely
  $\Delta h/h \sim 1,36h^2 \times 10^{-16}$.
  There exist , however, many other,
 more or less artificial, attempts to explain
 these flyby  anomalies \cite{Anders1}.

 \section{Discussion and Conclusion}

 In this letter we studied the implications of extending
  the gravitational energy shift for  photons    to all
 matter waves. The infinitesimal form of the
gravitational energy shift  implies  that the gravitational
attractive force  should in fact act between all mass-equivalent
energies.The corresponding space-time metric is found to be
exponential. A long time ago this metric was first  introduced by
Yilmaz \cite{Yilmaz} in an attempt to modify the Einstein's field
equations of general relativity. However, the Yilmaz theory was sharply
criticized\cite{Misner} on various grounds \cite{Ibison} as being
ill defined and without an event horizon occurrence
\cite{Robertson}. In our Lagrangian approach,
  the  exponential metric appears as one of the solutions
  leading to the observed gravitational energy shift. Neither
 use of general relativity nor the Yilmaz theory was made. In fact
 we used only
the theory of special relativity and the law of conservation of
total energy in the   binary system. Our particular solution
was based on the observational fact that a proper mass $m_0$ is
always observed as a gravitational mass $m_g = m_0e^u$ in a gravitational
field.  As a  consequence the
 gravitational attraction appears as
 acting between full mass-equivalent energies.

It is also well known that exponential metric  belongs to a large
class of   alternative theories of gravity, which all agree with the
solar observational tests. In contrast to   Einstein's general
relativity, the exponential metric does not predict formation of an
event horizon\cite{Robertson}.

We also tried to see whether a new force of gravity,
$F_g = F_Ne^u$ was able to explain  the Pioneer and flyby anomalies.
Unfortunately, the anomalies are too large  to be explained by
modifying the Newtonian gravity without violating  the well
established cartographic data  of the solar system .
Both, exponential and
Schwarzschild metrics give  unsatisfactory answers, $a =a_N(1 +2u +
...)$. However, both Pioneer spacecrafts and other flybys  have
changed their orbtal types from bound to unbound states
 and{\it vice versa} . bits into unbound ones.During this
maneuver the energy transfer took place violating the law of energy
conservation. Taking this into account,we produced  an explicit
formula by which the Pioneer anomaly can be explained. The same
mechanism works for flyby anomalies.

{\bf Acknowledgments}

 This work was supported by the Croatian Ministry of Science,
 Education and Sport, Project No.098-0982930-2900.

\end{document}